# Displacements and evolution of optical vortices in edge-diffracted Laguerre-Gaussian beams

Aleksandr Bekshaev[1*], Aleksey Chernykh[2], Anna Khoroshun[2], Lidiya Mikhaylovskaya[1]

[1]*Odessa I.I. Mechnikov National University, Dvorianska 2, 65082 Odessa, Ukraine*
[2]*East Ukrainian National University, Pr. Radiansky, 59-A, Severodonetsk, Ukraine*
[*]*Corresponding author: bekshaev@onu.edu.ua*

Based on the Kirchhoff-Fresnel approximation, we consider behavior of the optical vortices (OV) upon propagation of the diffracted Laguerre-Gaussian (LG) beams with topological charge $|m| = 1, 2$. Under conditions of weak diffraction perturbation (i.e. the diffraction obstacle covers only the far transverse periphery of the incident LG beam), the OVs describe almost perfect 3D spirals within the diffracted beam body, which is an impressive demonstration of the helical nature of an OV beam. The far-field OV positions within the diffracted beam cross section depend on the wavefront curvature of the incident OV beam so that the input wavefront curvature is transformed into the output azimuthal OV rotation. The results can be useful in the OV metrology and for the OV beam's diagnostics.

**1. Introduction**
During the past decades, light beams with optical vortices (OV) attract close attention of the optical community [1–4]. These intriguing optical objects, closely associated with the topological phase singularities, spectacularly illustrate the deep and fruitful optical-mechanical analogies and universality of physical laws [1–6] as well as provide a lot of inspiring applications in precise metrology [7–14], information transfer and processing [15–17] and micromanipulation techniques [18–20]. In particular, the edge diffraction of OVs has been studied intensively [21–33] enabling visual manifestation of the unique OV properties associated with their helical nature. One of the most impressive evidences of the helical energy flow in OV beams is the recently revealed spiral-like motion of the OV cores (amplitude zeros) within the diffracted beam cross section, that occurs when the screen edge performs a monotonous translation in the transverse direction towards or away from the beam axis [31–33].

The similar rotational motion of the OV cores is expected when the screen edge rests but the longitudinal evolution of the propagating diffracted beam is observed behind the screen. Under such conditions, the evolution of phase singularities in propagating edge-diffracted beams was studied in many details [28–30]. In these works, the main attention was paid to severely screened OV beams, in which the diffracted beam structure is strongly perturbed by the obstacle, and its evolution is accompanied by multiple topological events: the OV disappearance and regeneration, emergence of new OVs, their interactions and topological reactions, etc. This performance is interesting and valuable for diverse metrological purposes but the presence of additional factors mask the expected pattern of the OV migration and makes it not evident. Besides, in the mentioned works it was accepted that the diffraction screen is always situated in the OV beam waist cross section and, although the method for including an arbitrary wavefront curvature has been indicated, the diffraction of OV beams with a non-planar wavefront has not been analyzed nor discussed.

In the present paper, we fill this gap. Based on the simple model for the diffraction of circular Laguerre-Gaussian (LG) beams [25–30], we will consider the situations of weak diffraction perturbation (WDP) when the screen edge is located far enough from the incident beam axis. We do not specify exactly what perturbation can be called "weak". Practically this implies that the beam visually preserves the initial circular shape immediately after the screen, which for low-order LG beams takes place if the screen is distanced from the axis by two or more beam radii measured at $e^{-1}$ intensity level, but some important conclusions of the WDP-based reasoning can be applicable well beyond any formal limits of its validity [32]. Despite the model simplicity, we will show that the OV migration inside the propagating diffracted beam carries distinct and non-trivial fingerprints of the beam's helical nature. Moreover, the reaction of the OV positions on the incident wavefront curvature provides additional evidence for the beam rotational properties and additional means for their diagnostics.

## 2. General scheme and theoretical model of the LG beam diffraction

We assume the general scheme of the OV beam diffraction analyzed in [28–30] but the coordinate system is borrowed from [31–33] (see figure 1). If the incident beam in the screen plane S is described by the complex amplitude distribution $u(x_a, y_a)$, then in the observation plane at a distance $z$ behind S the diffracted beam complex amplitude can be found via the Kirchhoff-Fresnel integral

$$u(x,y,z) = \frac{k}{2\pi i z} \int_{-\infty}^{\infty} dy_a \int_{-\infty}^{a} dx_a\, u(x_a, y_a) \exp\left\{\frac{ik}{2z}\left[(x-x_a)^2 + (y-y_a)^2\right]\right\}. \tag{1}$$

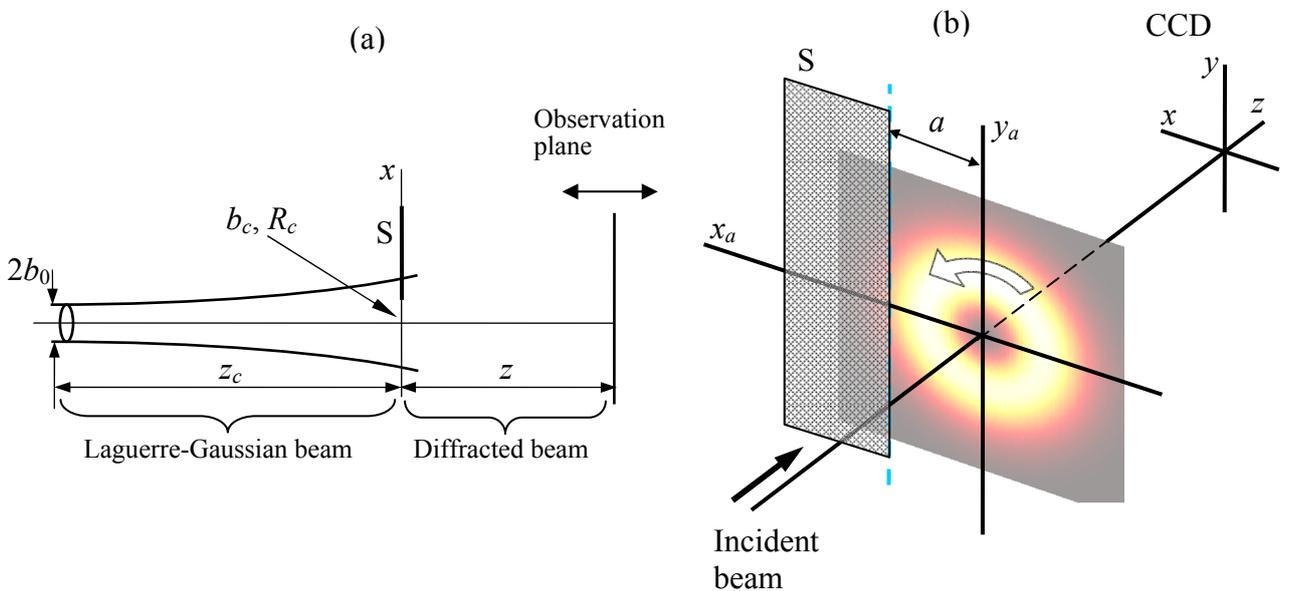

**Figure 1**. (a) Scheme of the incident LG beam diffraction; (b) magnified view of the beam screening and the involved coordinate frames. S is the diffraction obstacle (opaque screen with the edge parallel to axis *y*), the incident beam axis coincides with the axis *z*, the diffraction pattern is registered (e.g., by the CCD camera) in the observation plane whose position along *z* is adjustable. Further explanations see in text.

In our case the incident beam is the circular LG beam with zero radial index and topological charge $m$ for which $u(x_a, y_a) = u^{LG}(x_a, y_a, z_c)$ where [2,3,32]

$$u^{LG}(x_a, y_a, z_c) = \frac{(-i)^{|m|+1}}{\sqrt{|m|!}} \left(\frac{z_{Rc}}{z_c - iz_{Rc}}\right)^{|m|+1} \left(\frac{x_a + i\sigma y_a}{b_0}\right)^{|m|} \exp\left(\frac{ik}{2} \frac{x_a^2 + y_a^2}{z_c - iz_{Rc}}\right). \quad (2)$$

Here $\sigma = \text{sgn}(m) = \pm 1$, $k$ is the radiation wavenumber, $b_0$ is the Gaussian envelope waist radius, $z_c$ is the distance from the waist cross section to the screen plane (see figure 1(a)), and $z_{Rc} = kb_0^2$ is the corresponding Rayleigh length [34]; more usual "explicit" beam parameters, the current beam radius $b_c$ and the wavefront curvature radius $R_c$ in the screen plane, are related with these quantities by known equations

$$b_c^2 = b_0^2 \left(1 + \frac{z_c^2}{z_{Rc}^2}\right), \quad R_c = z_c + \frac{z_{Rc}^2}{z_c}. \quad (3)$$

For the LG beam (2), the Kirchhoff-Fresnel integral (1) can be calculated analytically, at least for $|m| = 1, 2, 3$ [25,28] but the known solutions are physically non-transparent, hardly interpretable, and it is our present task to expose some their important properties in an explicit form. To this end, we consider the case of a WDP, when the screen edge is placed at the far periphery of the incident beam cross section. In addition to numerical simulations, in this case we can employ the simplified analytical theory of the OV beam diffraction [32,33] which is outlined in the Appendix. It predicts that while $a \gg b_c$, the polar coordinates $r = \sqrt{x^2 + y^2}$ and $\phi = \arctan(y/x)$ of the OV core within the diffracted beam cross section obey the approximate relations

$$r = \left\{a^{|m|-1} \exp\left(-\frac{a^2}{2b_c^2}\right) \left|\frac{D_m(z)}{B_m(z)}\right|\right\}^{1/|m|}, \quad (4)$$

$$\phi = \frac{1}{m}\left[\arg D_m(z) - \arg B_m(z)\right] + \frac{ka^2}{2mR_c} + \frac{ka^2}{2mz} + \frac{2N}{m}\pi \quad (5)$$

where $N = 0, 1, \ldots |m|-1$, $D_m(z)$ and $B_m(z)$ are determined by equations (A10). Note that expression (5) describes the decomposition of a multicharged incident OV into the set of $|m|$ single-charged secondary OVs [30,32] numbered by $N$.

**3. OV displacements: incident beam with plane wavefront**

Following to works [28–30], we start with analyzing the single-charged incident LG beam with its waist at the screen plane, i.e. in equations (2) and (3)

$$z_c = 0, \quad b_c = b_0, \quad R_c = \infty. \quad (6)$$

Also, to preserve the symmetry and geometrical conditions of the previous works [31,33], we assume $m = -1$; the main beam parameters are taken as in [32],

$$b_c = 0.232 \text{ mm}, \quad k = 10^5 \text{ cm}^{-1}, \quad z_{Rc} = 53.8 \text{ cm}. \quad (7)$$

For a single-charged OV beam, the main consequence of the WDP is the OV displacement from its nominal position at the beam axis $x = 0$, $y = 0$. The positions of the OV cores at any distance $z$ behind the screen plane can be found numerically by the methods described elsewhere [30–32], and the results are given in figure 2 for four different screen-edge locations (see figure 1): $a = 3b_c$, $a = 2b_c$, $a = b_c$, and $a = 0.6b_c$. In order to abstract from the trivial component of the OV migration associated with the beam divergence, the transverse OV coordinates are normalized by the current beam radius

$$b = b_0 \sqrt{1 + \frac{z^2}{z_{Rc}^2}} \ . \tag{8}$$

As expected, the spiral-like trajectories are only observed under conditions of weak OV-beam perturbation ($a = 3b_c$ and $a = 2b_c$); the trajectories for $a = b_c$ and $a = 0.6b_c$ are given for the comparison (actually, these curves also contain certain spiral-like segments corresponding to the very small propagation distances, which are hardly available both to analysis and to observation because of the diffraction-fringes effects). On the other hand, under the WDP conditions, the OV displacement from the incident beam axis is rather small. This is clearly seen in figure 2(a) from confrontation of the blue spiral for $a = 2b_c$ against the red and black curves for $a = b_c$, and $a = 0.6b_c$; in case of $a = 3b_c$ the spiral is so small that it is separately magnified in figure 2(b). Note that the black curve in figure 2(a) corresponds to the curve marked "a = 0.6" in figure 5 of reference [29] but presents additional details due to the more favorable scale, and its orientation is changed because of the different coordinate frame and screen orientation.

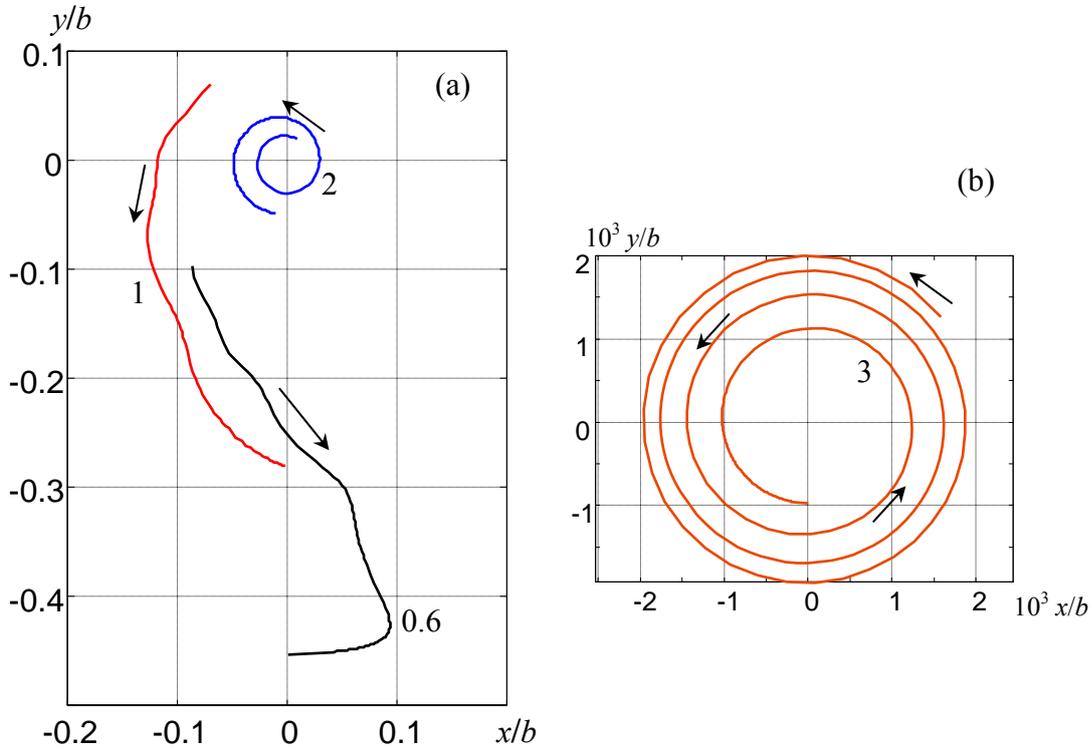

**Figure 2**. Evolution of the OV core position in the cross section of the diffracted LG beam of equations (2), (6) with $m = -1$ and $R_c = \infty$ for (a) $a = 2b_c$ (blue), $a = b_c$ (red), $a = 0.6b_c$ (black) and (b) $a = 3b_c$ (brown); each curve is marked by the corresponding value of $a$ in units $b_c$. Arrows indicate directions of the OV motion, initial points of the curves correspond to $z = 10$ cm $= 0.186z_{Rc}$ (near field), final points correspond to $z = \infty$ (far field).

The 3D behaviour of the spirals for $a = 2b_c$ and $a = 3b_c$ in the propagating diffracted beam is illustrated by figure 3. Here the longitudinal coordinate $z$ varies within the range

$$10 \text{ cm} = 0.186z_{Rc} < z < 600 \text{ cm} \approx 11.15z_{Rc} \ .$$

Note that the OV core motion along the spiral trajectory is not uniform: almost the whole observable evolution happens at first 10% of the full range of $z$ variation (cf. figure 4). To present the 3D trajectories more conveniently, in figures 3(a), 3(b) the longitudinal scales are non-uniform: the real distance $z$ is normalized by the scale factor (8). Remarkably, even with this precaution, the region $z < 10$ cm cannot be shown properly because with reducing $z$, the rate of the spiral rotation (theoretically) infinitely grows while the spiral pitch (distance between consecutive coils) infinitely decreases. This is supported by the simplified asymptotic model (5) which leads to the approximate rule

$$\phi \approx \text{const} + k\frac{a^2}{2mz}. \tag{9}$$

Obviously, with growing $z$, the rotation practically stops whereas at small $z$ the rate of rotation $d\phi/dz$ can be rather high. However, this implies no unphysical divergence for very small $z$ because the paraxial form of the Kirchhoff-Fresnel integral (1) and all the ensuing considerations of this paper are only valid for $z \gtrsim 2\cdot 10^3 k^{-1}$ [35], which in our present conditions means $z \gtrsim 2\cdot 10^{-2}$ cm. Moreover, when $z \to 0$, the relation $D_m(z) \to 0$ also takes place (see (A10)), i.e. with growing rate of rotation, the OV off-axial displacement (4) becomes so small that the "theoretic" spiral motion is practically imperceptible.

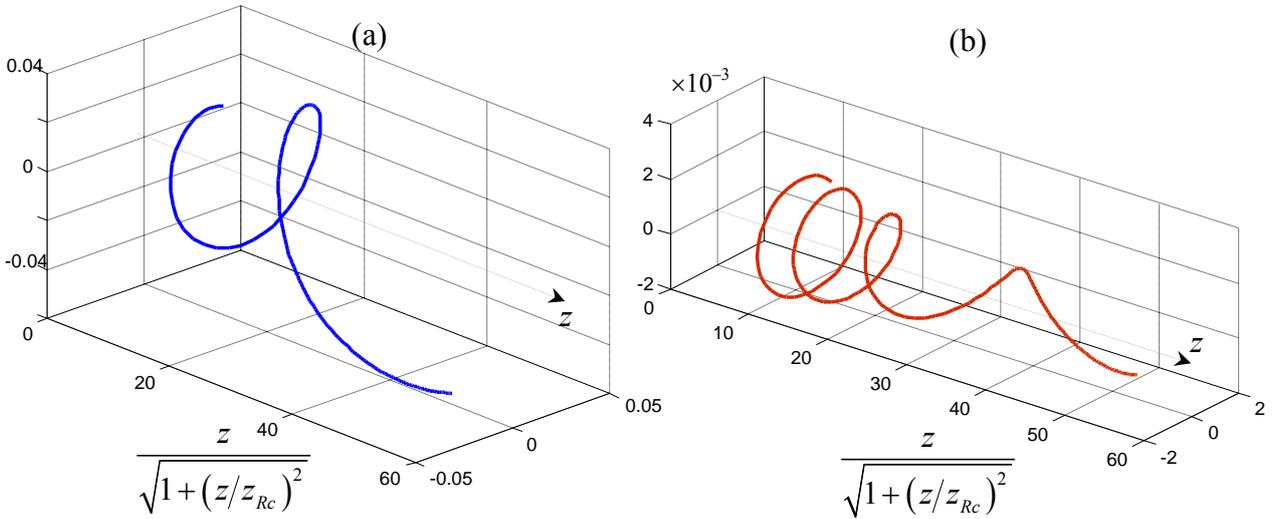

**Figure 3**. 3D evolution of the OV core positions in the diffracted LG beam (2), (6) with $m = -1$ and $R_c = \infty$ for (a) $a = 2b_c$ and (b) $a = 3b_c$. The blue and brown spirals in figure 2 represent the projections of these trajectories viewed against axis $z$.

Importantly, at $z \to \infty$ all the OV trajectories in figure 2 approach the vertical axis. This fact was noticed previously and interpreted basing on the general concept that the diffracted beam intensity distribution rotates during propagation [29,30]. Just after the screen, the diffracted beam intensity distribution looses the circular symmetry but preserves the mirror symmetry with respect to an axis orthogonal to the screen edge. In the course of further propagation, the beam shape continuously changes, generally, in a rather complicated way. But this transformation possesses a sort of regular rotational component, which, in the far field, results in the intensity distribution with another symmetry axis, orthogonal to the initial one and parallel to the screen edge. Accordingly, in the far field, the OV of the diffracted beam eventually approach this symmetry axis that in normalized coordinates of figure 2 coincides with the $y$-axis.

### 4. OV displacements: incident beam with spherical wavefront

The last conclusion of the previous section was drawn from the OV diffraction analysis based on an assumption that the wavefront of the incident OV beam is plane [29,30]. Now consider what influence can be caused by the wavefront curvature. To this end, it would be suitable to involve the simple rule for transformation of the diffracted beam complex amplitude that occurs if the incident beam is modified according the equation

$$u(x_a, y_a) \to u(x_a, y_a) \exp\left( ik \frac{x_a^2 + y_a^2}{2R_c} \right) \tag{10}$$

(e.g., the plane front is replaced by the spherical one with the same intensity profile) [28]. In this situation, if the initial distribution $u(x_a, y_a)$ produces the diffracted beam with complex amplitude $u(x, y, z)$, the modified initial beam (10) produces the diffracted beam distribution

$$u_e(x, y, z) = \frac{1}{1 + \frac{z}{R_c}} \exp\left[ ik \frac{x^2 + y^2}{2(z + R_c)} \right] u\left( \frac{x}{1 + z/R_c}, \frac{y}{1 + z/R_c}, \frac{z}{1 + z/R_c} \right). \tag{11}$$

To elucidate the meaning of this rule let us suppose that the incident beam $u(x_a, y_a)$ possesses the plane wavefront (except the helical component associated with the term $(x_a + i\sigma y_a)^{|m|}$ in (2)). Also, we suppose that this beam ('prototype beam'), being diffracted, produces in the cross section $z_0$ behind the screen such complex amplitude distribution for which the OV core is situated in the point $(x = x_0, y = y_0)$. Hence, (11) dictates that the modified incident beam (10) produces the diffracted beam in whose cross section

$$z = \frac{z_0}{1 - z_0/R_c} \tag{12}$$

the OV core is located in the point

$$x_r = \frac{x_0}{1 - z_0/R_c}, \quad y_r = \frac{y_0}{1 - z_0/R_c}. \tag{13}$$

In particular, the far field of the modified spherical-front beam (10) is realized when $z \to \infty$ which corresponds to the finite cross section of the prototype beam, $z_0 = R_c$. Note that transformation (13) affects only the off-axial distance of the OV position while its azimuth $\phi_r = \arctan(y_0/x_0) = \arctan(y_r/x_r)$ remains the same as in the prototype plane-front beam.

This reasoning suggests the simple procedure for determining the azimuthal far-field positions of the OV cores in the diffracted LG beam with non-planar wavefront. It requires the OV trajectory for the prototype plane-front beam (see figure 4 where the blue spiral of figure 2(a) is magnified and furnished with marks denoting the propagation distances behind the screen in centimetres). For example, when the incident beam wavefront possesses the curvature radius $R_c = 80$ cm, equation (12) shows that the far field for the diffracted beam is realized at $z_0 = R_c = 80$ cm. Accordingly, the corresponding OV azimuth coincides with the azimuth of the red rectangle marked "80". And indeed, the independently calculated OV trajectory for the incident beam (2) with $R_c = 80$ cm (green curve) is oriented close to this direction, and the small discrepancy can be explained by the limited range of the propagation distances accepted for the green curve calculation (the "genuine" far field with infinite propagation distance was never reached). The same is correct for other examples corresponding to $z_0 = R_c = 18$ cm (red curve), 30 cm (brown curve) and infinity (black curve). This reasoning distinctly shows that the far-field OV position belongs to the symmetry axis parallel to the screen edge only if the incident LG beam possesses a plane wavefront.

Further application of this procedure to cases with $R_c < 0$ (converging beams) requires the knowledge of the plane-front diffracted beam behaviour at $z_0 < 0$, which seems non-physical. However, the Kirchhoff-Fresnel integral (1) formally is valid for any $z$; moreover, equations (1) – (3) suggest that

$$u^*(x_0, y_0, -z_0) = u(x_0, -y_0, z_0)$$

where the asterisk denotes the complex conjugation. This means that positions of the amplitude zeros at negative $z_0$ can be easily found once they are known for positive $z_0$. Additionally, to get the far-field OV positions, we should take into account that in (12) $z$ tends to positive infinity, and for negative $z_0$ this implies that transformation (13) inverts the signs of both transverse coordinates. As a result, the 'prototype beam' OV trajectory for negative $z_0$ can be obtained by a mirror reflection of the blue spiral described above, and is presented in figure 4 as a pale grey dashed curve. It can be employed exactly in the same manner as the blue curve itself, with the help of corresponding distance marks some of which are explicitly indicated.

Whereas for diverging beams (positive $R_c$) the symmetrical structure of the intensity distribution, typical for the plane-front beams, does not exist in the "physical" range $z > 0$ (the far-field intensity pattern reproduces the prototype beam structure at finite $z_0$), for converging beams with $R_c < 0$ such a possibility is realized at finite $z$. According to (12) and (13), this occurs when $z = -R_c > 0$, which corresponds to $z_0 = \infty$. It is expectable because the plane $z = -R_c > 0$ is actually a focal plane of the 'lens' performing transformation (10), and in this plane the Fraunhofer diffraction takes place that is equivalent to the far field propagation [34].

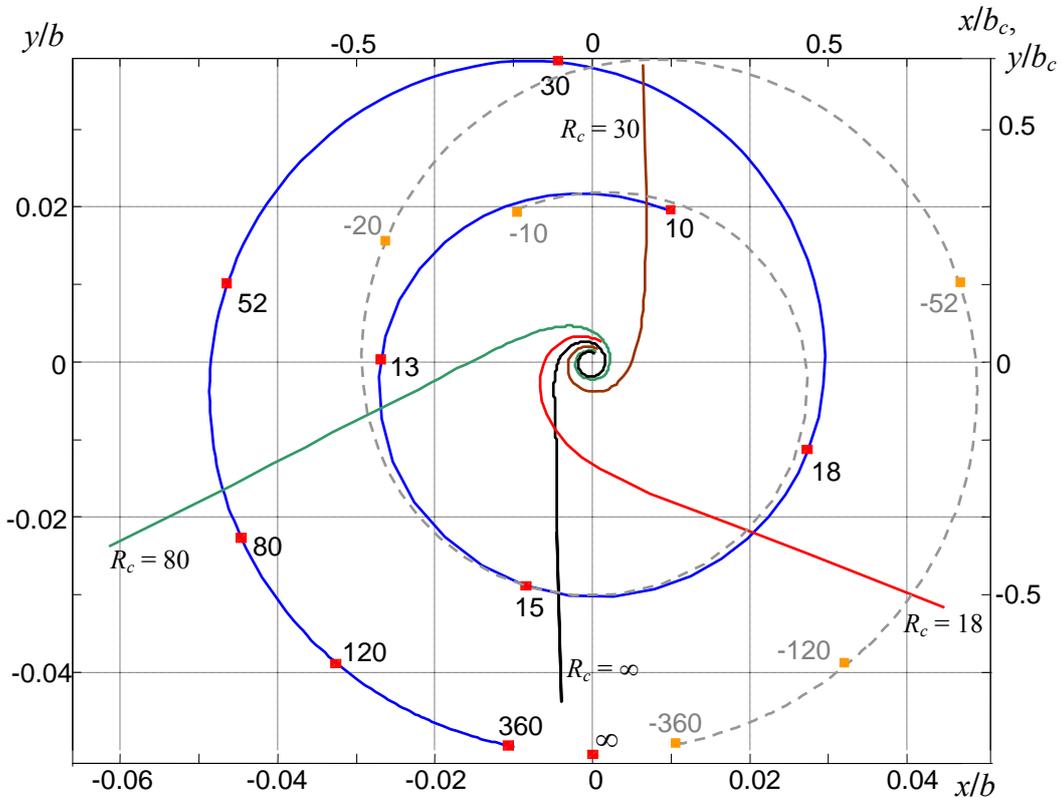

**Figure 4**. Illustration of the far-field azimuthal OV positions in diffracted beams with non-planar wavefront for $m = -1$ and $a = 2b_c$. Blue curve is the transverse projection of the OV trajectory in the prototype plane-front diffracted beam (cf. figure 2(a), blue curve, and figure 3(a)), pale grey dotted curve represents its branch for the negative propagation distances. The left and bottom scales show the current OV coordinates in normalized units (8), red (yellow) rectangles denote the propagation distance marked in centimeters. Red, brown, green and black curves are the transverse projections of the OV trajectories in the modified spherical-front diffracted beams (10) (radius of curvature is indicated near each curve, the OV coordinates are measured in units of $b_c$, and marked in the right and top scales).

The resulting azimuthal OV positions in diffracted LG beams with spherical wavefronts are illustrated in figure 5. Note that when $R_c \to \infty$, all the curves approach the azimuth values $\phi_r = (3/2)\pi$, $(7/2)\pi$ and $(15/2)\pi$, that means that the OV trajectories always end at the negative half-axis $y$, as was discussed above. This example also shows advantages of the WDP conditions. Theoretically, in cases of strong beam screening, the wavefront curvature also affects the OV positions in the diffracted beam cross section but the red and black curves in figure 5 display a rather limited range of possible azimuthal deviations, and their sensitivity to the incident wavefront curvature appears to be low.

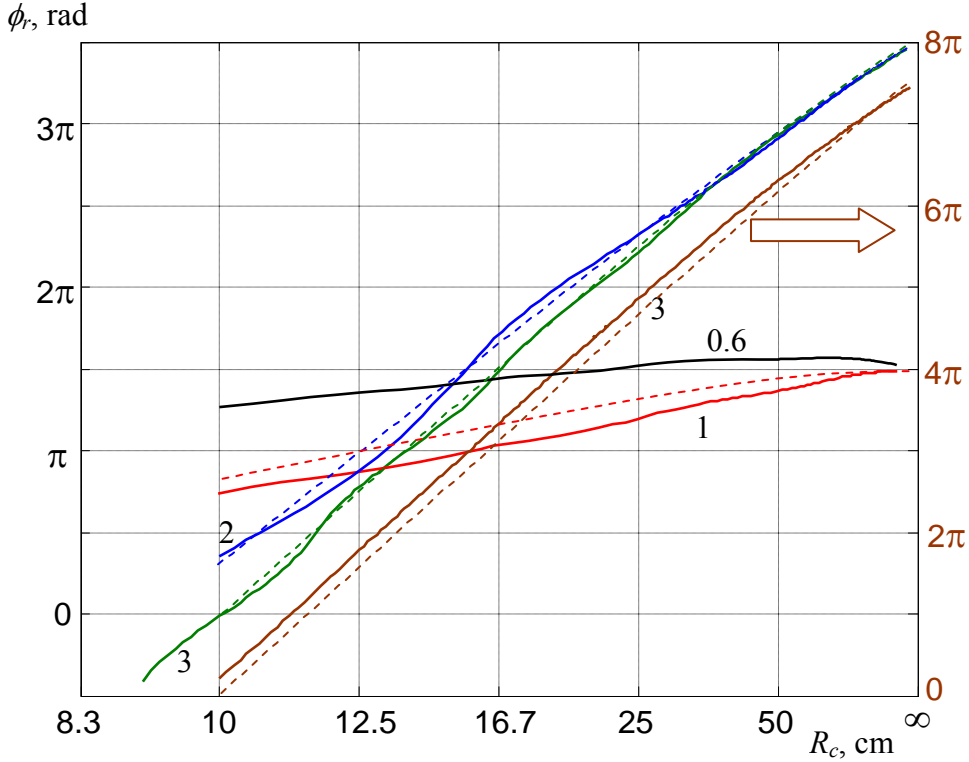

**Figure 5**. Far-field azimuthal OV positions in the diffracted LG beam (2) with $m = -1$ vs the wavefront curvature radius: (brown curve) $a = 3b_c$, right vertical scale; (blue curve) $a = 2b_c$ (cf. figure 4); (red curve) $a = b_c$ (cf. figure 2(a)); (black curve) $a = 0.6b_c$ (cf. figure 2(a)). The green curve represents the far-field orientation of the straight line connecting two secondary OVs in the diffracted LG beam (2) with $m = -2$, $a = 3b_c$ (cf. figure 7, dashed line). Each curve is marked by the corresponding value of $a$ in units $b_c$; dashed lines show analytical approximations (14) for the solid curves of the same colors.

All these conclusions are based on the numerical simulations but they are also justified by the approximate model of (4), (5). First, we note that if we introduce a finite $R_c$ by means of transformation (10), the set of parameters (6) becomes modified: $b_c$ remains the same as in (7) but it is no longer equal to the new waist radius $b_0$, and $z_c$ is determined by the new waist position. Accordingly, the Rayleigh length of the modified LG beam $z_{Rc} = kb_0^2$ differs from the value of (7) but equations (3) entail the relation

$$\frac{z_c}{z_{Rc}} = \frac{kb_c^2}{R_c}.$$

Then, with the help of equations (A10) one can easily derive the far-field ($z \to \infty$) representation of the OV azimuth (5):

$$\phi_\infty = \frac{|m|}{m}\left(\frac{\pi}{2} - \arctan\frac{kb_c^2}{R_c}\right) + \frac{ka^2}{2mR_c} + \frac{1}{2m}\arctan\frac{kb_c^2}{R_c} + \frac{2N}{m}\pi. \quad (14)$$

The corresponding dependences $\phi_\infty(R_c)$ for $m = -1$ ($N = 0$) are presented in figure 5 by the brown, blue and red dashed curves (since the azimuth values differing by a complete angle are equivalent, upon constructing the dashed curves necessary integer numbers of $2\pi$ were added to the immediate results of (14)). One can see that expression (14) provides a rather good approximation for the precise numerical data, especially at large $R_c$ and $a = 2b_c$ and $a = 3b_c$; when the screening grows (i.e., for $a = b_c$) the approximation looses its quantitative accuracy but remains qualitatively valid.

## 5. Incident beam with second-order OV

The similar WDP-induced behavior is typical for higher-order OV beams. As an example, we consider the diffraction of a charge-2 LG beam (2) with $m = -2$ and the same parameters (6), (7). The main difference from the charge-1 case is that the diffraction causes the incident OV decomposition into two single-charged secondary OVs that evolve separately within the diffracted beam "body". At the WDP conditions, they form a "double spiral" (figure 6), each component being quite similar to the OV trajectories observed in the diffracted beam with $|m| = 1$ (figure 3). Transverse projections of the trajectories shown in figure 6 are presented in more detail in figure 7; similarly to figure 4, the distance marks are added denoting the current longitudinal positions. Just as in figure 4, the distance marks can be used for prediction of the azimuthal orientation of the far-field OV displacement if the incident beam wavefront is not plane (see the black, green and cyan curves for $R_c = 50$ cm, $R_c = 85$ cm and $R_c = \infty$, correspondingly).

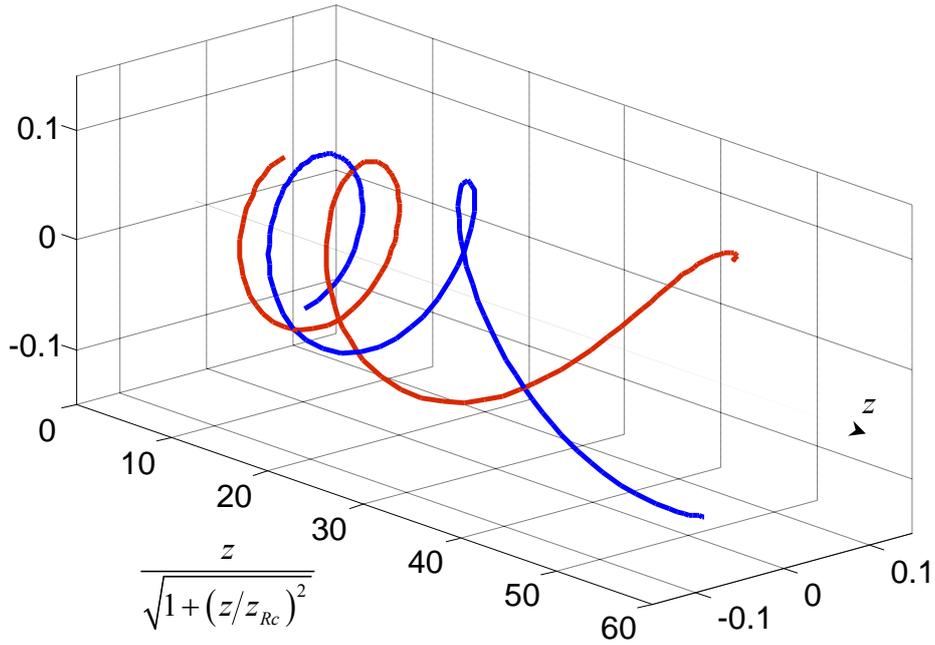

**Figure 6**. 3D evolution of the OV core positions in the diffracted LG beam (2), (6) with $m = -2$ and $R_c = \infty$ for $a = 3b_c$. As in figure 3, the transverse coordinates are in units of $b$ (8), and the longitudinal coordinate is normalized by the scale factor of (8).

Note that the two secondary OVs are always situated oppositely with respect to the nominal beam axis $z$. However, they never form a perfect central-symmetric pair (with the centre at the transverse coordinate origin) dictated by the approximate model (5), except in the asymptotic case $z = \infty$, $R_c = \infty$; this is also a consequence of the slight symmetry perturbation upon the WDP. At the same time, the two secondary OVs define a straight line whose far-field orientation can be distinctly associated with the wavefront curvature of the incident LG beam (see, for example, the dashed line $L$ in figure 7 that unites the two OV cores in case $R_c = 50$ cm). The rotation of this line with the incident wavefront curvature variation is illustrated by the green curve in figure 5. As in case of $|m| = 1$, its behaviour can be fairly approximated by equation (14) (see the green dashed curve in figure 5). Its main difference from the simulation results is that it does not show the rotation accelerations near $\phi_r = \pi/2, 3\pi/2$ where one of the OVs crosses the negative $y$ half-axis. These accelerations correspond to the trajectories' "jumps" observed in the diffracted Kummer OV beams and can be explained within the frame of the improved analytical model for the OV beam diffraction [32,33,36].

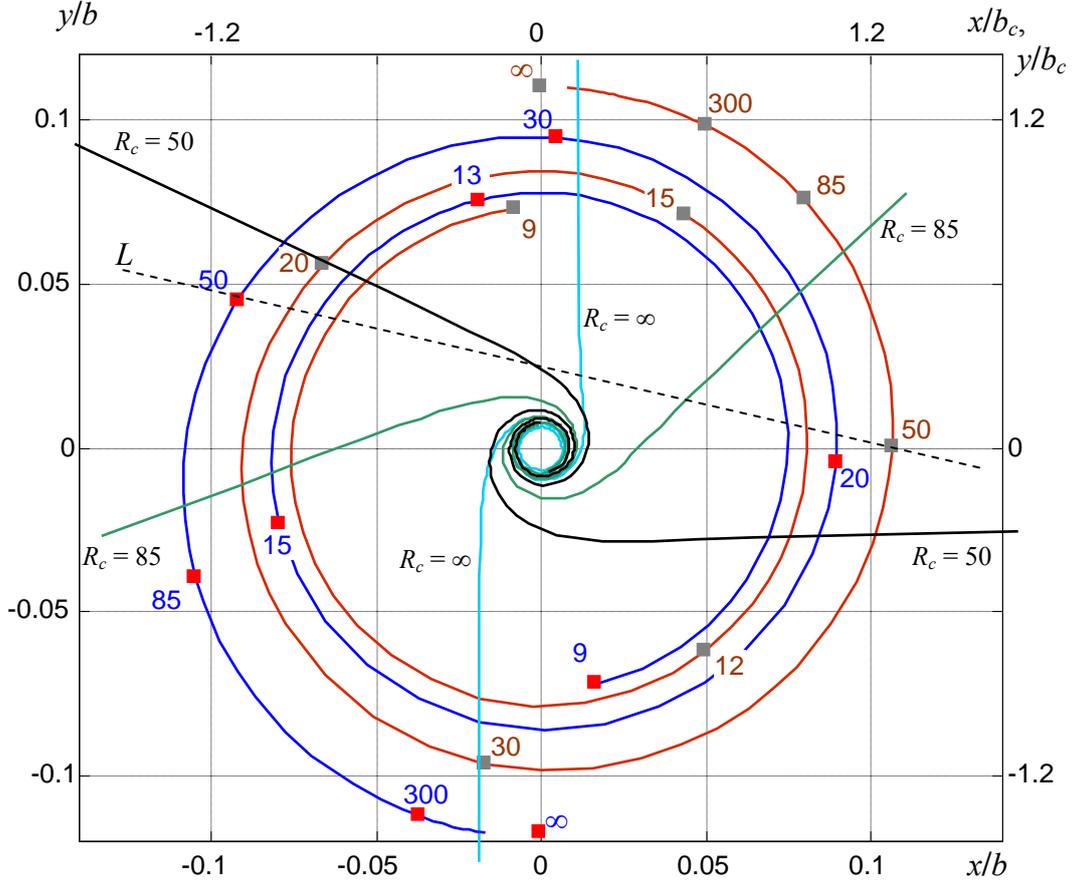

**Figure 7**. Blue and brown curves represent trajectories of the two secondary OVs formed in the diffracted LG beam cross section with $m = -2$ and $a = 3b_c$ (cf. figure 6), red (grey) rectangles denote the propagation distance (longitudinal coordinate) marked in centimeters, left and bottom scales show the transverse OV coordinates in normalized units (8). Black, green and cyan curves are the transverse projections of the secondary-OV trajectories in the modified spherical-front diffracted beams (10) (radius of curvature is indicated near each curve, the OV coordinates are measured in units of $b_c$, and marked in the right and top scales). Dashed line $L$ shows the far-field orientation of the OV pair at $R_c = 50$ cm.

In the whole, the far-field pattern of the diffracted OV beam supplies an interesting example where the input wavefront curvature is transformed into the output azimuthal rotation of the secondary OV pair, which can possibly find applications for the wavefront diagnostics and measurements.

## 6. Conclusion

Now let us summarize the main outcome of this paper. Main results presented above confirm that the simple and ubiquitous situation of edge diffraction provide additional impressive manifestations of the helical nature of light beams with OV. The diffraction obstacle introduces the beam perturbation that causes the OV displacement from its original axial position (for an *m*-charged OV, |*m*| single charged displaced OVs are formed around the axis). Then, while the diffracted beam freely propagate, the displaced OVs migrate over the beam cross section along spiral trajectories, initially (at small post-screen distances) with high rotation rate, which rapidly decreases and practically stops far enough behind the screen, in compliance with the approximate relation (9). The most articulate spiral-like trajectories occur under conditions of weak diffraction perturbation (WDP), when the screen edge is separated from the incident beam axis by two or more beam radii, i.e. when the beam circular shape is almost unaffected by the diffraction. This puts certain limitations to the practical employment of the predicted behavior, since the off-axial OV displacements are small and their measurement can meet difficulties. However, we emphasize the principal importance of the predicted features and their close relation with the physical nature of OVs; besides, at least for a multicharged OV diffraction, even under the WDP conditions the typical OV displacement can reach ~10% of the current beam radius *b* (see figures 5, 6), which is quite measurable.

Additional interesting possibilities are associated with relations between the incident beam wavefront curvature and the diffracted beam structure. First, we have refined the earlier statements [29,30] that in the far field, the diffracted circular OV beam acquires the symmetry with respect to an axis parallel to the screen edge. In fact, this symmetry is only realized in the beam Fraunhofer (Fourier) plane, which can be real (if the beam converges, radius of curvature $R_c < 0$) or imaginary (if $R_c > 0$), and only for beams with plane wavefront it occurs at the propagation infinity (in the far field). Importantly, the actual far-field pattern of the diffracted beam essentially depends on the incident beam wavefront curvature. In terms of the OV displacements, any change of the incident beam wavefront curvature, with preserving the intensity profile, induces the azimuthal rotation of the far-field OV position in the diffracted beam. This sensitivity may be useful for the OV beam diagnostic and for the wavefront measurements.

It should be mentioned that the same information about the incident beam parameters and the diffraction conditions can be extracted from the diffracted beam reaction on the "strong" diffraction perturbation, i.e. when the screen – axis distance is comparable or less than the beam radius, $a \leq b_c$. However, in this case the information is "hidden" in the whole diffracted beam structure and there are no distinctly visible details that can be directly associated with the specific characteristics of the incident beam. For example, although the red and black curves of figure 2(a) demonstrate much higher (and, expectably, more suitable for measurements) OV displacements, there is no distinct rotation that could evidence for the beam helical structure. Besides, whereas the rule for finding the far-field azimuthal OV position described in figure 4 still works for the OV trajectories in strongly screened spherical-wavefront beams, the range of possible azimuthal deviations of the far-field OV is rather limited, which results in their low sensitivity to the incident wavefront curvature (cf. red and black curves in figure 5).

In this paper we did not touch an interesting aspect of the far-field OV rotation which associates the diffracted beam behaviour with the Gouy phase of the incident LG beam (see, e.g., [27,30]. We hope that the present approach will be helpful in further studies of this aspect, including the vector beams demonstrating its impressive manifestations [37–39].

Finally, we note that the edge diffraction is just one example of the OV beam transformation with symmetry violation. There are other such transformations; in particular, the astigmatic focusing and telescopic transformations of OVs are studied in much detail (see, e.g., [40,41]). But they show quite different behaviour of the secondary OVs within the transformed beam cross section, most likely due to preserving the central symmetry of the beam transverse profile.

However, one can expect that some features of the OV migration, similar to those considered in this paper, can be detected in OV beams subjected to the symmetry violation which destroys the central symmetry. It would be meaningful and instructive to inspect the response of an OV beam to the unilateral beam constraint, e.g., by a "soft" diaphragm, or even unilateral phase modification in some peripheral part of the beam cross section. These transformations are, properly, special cases of diffraction but they admit more 'gentle' conditions suppressing the edge waves, diffraction fringes [34], etc., which will be favourable for precise measurements. This is important because, again, the most demonstrative behaviour is expected under the WDP conditions where the modifications of the OV beam structure (e.g., the OV displacements, etc.) are small.

**Appendix**

Equation (1) can be recast in the form

$$u(x,y,z) = u^I(x,y,z) - \frac{k}{2\pi i z}\int_{-\infty}^{\infty} dy_a \int_a^{\infty} dx_a\, u(x_a, y_a) \exp\left\{\frac{ik}{2z}\left[(x-x_a)^2 + (y-y_a)^2\right]\right\} \quad (A1)$$

where $u^I(x,y,z)$ is the complex amplitude distribution of an unperturbed incident beam that would take place in the observation plane if the screen were absent. This form is suitable for asymptotic analysis of the integral in case of WQP, when $a \gg b_c$.

In our case of the incident LG beam, $u(x_a, y_a)$ is expressed by (2), and $u^I(x,y,z) = u^{LG}(x,y,z_c + z)$. Then, omitting the coordinate-independent multipliers of (2), the integral in (A1) can be written in the form

$$\exp\left[\frac{ik}{2z}(x^2 + y^2)\right]\int_{-\infty}^{\infty} dy_a\, P(y_a, y, z_c - iz_{Rc}) \int_a^{\infty} dx_a\, (x_a + i\sigma y_a)^{|m|} P(x_a, x, z_c - iz_{Rc}) \quad (A2)$$

where

$$P(x_a, x, d) = \exp\left[\frac{ikx_a^2}{2}\left(\frac{1}{d} + \frac{1}{z}\right) - \frac{ik}{z}xx_a\right]. \quad (A3)$$

In further calculations, we seek the positions of the OV cores which, under the WDP conditions, are close to the beam axis, so we can assume

$$x \simeq y \simeq 0 \quad (A4)$$

and neglect the summands proportional to $x$, $x^2$ and $y$, $y^2$ when compared to the coordinate-independent terms. Additionally, for large $a$, the internal integral in (A2) can be estimated via the asymptotic relation valid for arbitrary function $f(x)$:

$$\int_a^{\infty} f(x)\exp(iKx^2) dx \simeq \frac{i}{2K}\frac{f(a)}{a}\exp(iKa^2) + O\left(\frac{1}{a^2}\right)$$

(it can be derived due to integration by parts). As a result, we get

$$\int_a^{\infty} dx_a\, (x_a + i\sigma y_a)^{|m|} P(x_a, x, z_c - iz_{Rc}) \simeq \frac{iP(a, 0, z_c - iz_{Rc})}{ak\left(\frac{1}{z_c - iz_{Rc}} + \frac{1}{z}\right)}(a + i\sigma y_a)^{|m|}. \quad (A5)$$

In turn, the external integral of (A2) is estimated by the method of stationary phase which in couple with condition (A4) gives

$$\int_{-\infty}^{\infty} dy_a\, (a + i\sigma y_a)^{|m|} P(y_a, y, z_c - iz_{Rc}) \simeq \sqrt{\frac{2\pi i}{k\left(\frac{1}{z_c - iz_{Rc}} + \frac{1}{z}\right)}}\, a^{|m|}. \quad (A6)$$

Hence, restoring the coordinate-independent multipliers of (2) omitted in (A2), we obtain the asymptotic representation for the integral term in the right-hand side of (A1):

$$-\frac{(-i)^{|m|+1}}{\sqrt{|m|!}}\sqrt{\frac{i}{2\pi}}\frac{k}{z}\left(\frac{z_{Rc}}{z_c - iz_{Rc}}\right)^{|m|+1}\left[k\left(\frac{1}{z}+\frac{1}{z_c - iz_{Rc}}\right)\right]^{-3/2}\frac{a^{|m|-1}}{b_0^{|m|}}P(a,0,z_c - iz_{Rc}). \quad (A7)$$

The first term of (A1), with allowance for the near-axis condition (A4), reads

$$u^I(x,y,z) = u^{LG}(x,y,z_c + z) \simeq \frac{(-i)^{|m|+1}}{\sqrt{|m|!}}\left(\frac{z_{Rc}}{z_c + z - iz_{Rc}}\right)^{|m|+1}\left(\frac{x + i\sigma y}{b_0}\right)^{|m|}. \quad (A8)$$

Now, combining this with the integral term (A7) and accounting for (A3), we arrive at the final asymptotic representation of the diffracted beam complex amplitude distribution

$$u(x,y,z) \simeq \frac{b_0}{\sqrt{|m|!}}\left(-\frac{iz_{Rc}}{b_0}\right)^{|m|+1}\left\{B_m r^{|m|}\exp(im\phi) - D_m a^{|m|-1}\exp\left[\frac{ika^2}{2}\left(\frac{1}{z_c - iz_{Rc}}+\frac{1}{z}\right)\right]\right\} \quad (A9)$$

where

$$B_m(z) = \frac{1}{(z + z_c - iz_{Rc})^{|m|+1}}, \quad D_m(z) = \sqrt{\frac{i}{2\pi}}\frac{k}{z}(z_c - iz_{Rc})^{-|m|-1}\left[k\left(\frac{1}{z}+\frac{1}{z_c - iz_{Rc}}\right)\right]^{-3/2}. \quad (A10)$$

Note that (A9) can be interpreted as a superposition of the unperturbed incident beam and the edge wave apparently "emitted" by the screen edge [34]. Hence, the coordinates of the OV cores can be found as zeros of expression (A9), which with account for (3) leads to equations (4) and (5).